\documentclass[aps,prx,twocolumn, superscriptaddress,amsmath,amssymb,longbibliography]{revtex4-2}
\usepackage{float}
\usepackage[english]{babel}
\usepackage[utf8]{inputenc}
\usepackage{siunitx}
\usepackage[colorinlistoftodos, color=green!40, prependcaption]{todonotes}
\begin{document}
\preprint{APS/123-QED}
\title{Probabilistic denoising for reliable signal extraction in spectroscopy}
\author{Younsik Kim}
     \email[]{leblang@snu.ac.kr}

    \affiliation{Department of Physics and Astronomy, Seoul National University, Seoul, Korea}
\author{Changyoung Kim}
     \email[]{changyoung@snu.ac.kr}
    \affiliation{Department of Physics and Astronomy, Seoul National University, Seoul, Korea}
\date{\today} 

\begin{abstract}
While deep learning offers powerful capabilities for scientific research, its application is often hindered by a lack of quantitative reliability. To address this, we introduce a probabilistic denoising framework that simultaneously extracts denoised signals and element-wise predictive uncertainties from noisy data. We demonstrate this approach on three-dimensional angle-resolved photoemission spectroscopy data, showing that the model reliably recovers the spectral features of a cuprate superconductor from Poisson-distributed noise with an average count of only 0.02 electrons per voxel. Crucially, we show that these predicted uncertainties can be propagated into subsequent superconducting gap analyses, enabling quantitative parameter extraction with scientifically meaningful error bars. Furthermore, we validate the broad applicability of our approach by successfully extending it to two-dimensional X-ray diffraction data. Ultimately, this approach establishes uncertainty-aware deep learning not merely as a visualization tool, but as a rigorous framework for scientific data analysis.

\end{abstract}


\maketitle

\begin{figure*}[!t]
	\centering
	\includegraphics[width=1\textwidth]{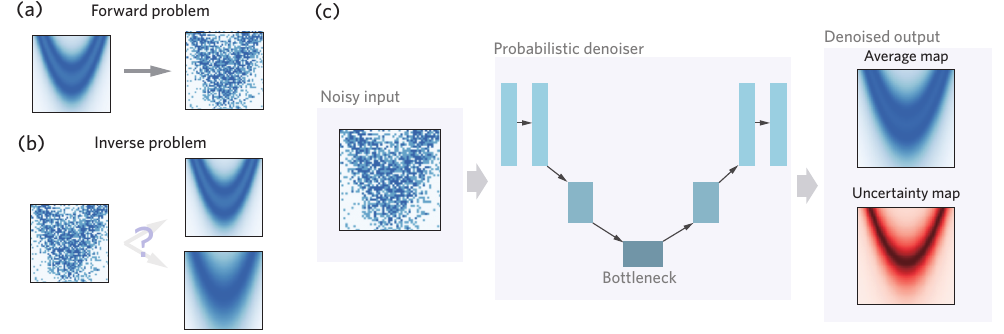}
	\caption{
         {\bf Overview of probabilistic denoising for spectroscopic data with uncertainty quantification.}
         (a) Forward problem: adding noise to clean spectra (well-defined).
         (b) Inverse problem: recovering the original spectrum from noisy data (ill-posed due to information loss).
         (c) Probabilistic denoiser: a neural network maps a noisy input to a denoised output (predictive mean) and an uncertainty map (predictive standard deviation), providing voxel-wise confidence estimates.
         }
	\label{Fig1}
\end{figure*}

Despite its remarkable capabilities in solving complex nonlinear problems, why has the integration of deep learning into scientific research been notably slower than in other practical domains? One of the answers lies in the fundamental gap between empirical performance and scientific rigor~\cite{roscher2020explainable,abdar2021review,karniadakis2021physics}. While industrial applications often prioritize performance, scientific applications require reliability and interpretability of the results. Standard deep learning models, often criticized as "black boxes", typically provide point estimates without a clear representation of the underlying uncertainty and intermediate reasoning~\cite{abdar2021review}.

These issues are particularly critical in signal restoration tasks, such as denoising. While corrupting clean data with noise is a well-defined forward problem (Fig. 1(a)), the recovery of the original clean data from noisy data is an ill-posed inverse problem~\cite{tarantola2005inverse,bertero1988ill,kendall2017uncertainties}. Since noisy data inherently carry imperfect information, removing noise and thus restoring the original clean signal is highly ambiguous, which can result in multiple plausible solutions (Fig. 1(b)). This inherent ambiguity in denoising hinders reliable analysis and necessitates the quantification of statistical uncertainty for successful application in scientific domains. Indeed, uncertainty-aware deep learning has attracted growing attention as a promising direction to meet this need~\cite{gawlikowski2023survey,psaros2023uncertainty,ghosh2021ensemble,vasudevan2021off}.

One natural realization of this idea is to adopt a probabilistic framework that formulates denoising as the estimation of the underlying probability distribution of the signal. In such an approach, the model simultaneously predicts the signal (mean) and its associated uncertainty (variance), by optimizing a negative log-likelihood (NLL) objective function~\cite{kendall2017uncertainties}. However, implementing this framework in deep learning models remains challenging. Training a network to produce denoised output with uncertainty estimates is often unstable requiring delicate optimizations. Consequently, despite the well-established theoretical foundation, probabilistic denoising has rarely been applied to spectroscopic data analysis. 

In this work, we realize this probabilistic approach for three-dimensional (3D) angle-resolved photoemission spectroscopy (ARPES). The network is successfully trained to simultaneously provide the denoised signal and element-wise (voxel-wise for the 3D data considered here) predictive variance from extremely noisy data following a Poisson distribution. Despite an average count of only 0.02 electrons per voxel, the model reliably denoises the signal while providing uncertainty estimates for each voxel. We further demonstrate that these voxel-wise intensity uncertainties can be propagated into subsequent superconducting gap analyses, enabling quantitative parameter extraction with scientifically meaningful error bars. To highlight the broad applicability of our approach, we also apply this method to two-dimensional (2D) X-ray diffraction (XRD) data. This approach establishes uncertainty-aware deep learning as a practical framework for early estimation of physical parameters without compromising scientific rigor.


\section*{Results}

To achieve high performance in denoising 3D ARPES data, we adopted an autoencoder denoising architecture~\cite{masci2011stacked}. This architecture ensures a large receptive field while maintaining a relatively compact network size (see Supplemental Material Section~1 for the details of the network~\cite{supple}). Such a feature is essential for 3D data, as the network would otherwise become unmanageably large for typical graphics processing units (GPUs). The receptive field of the network used in this work is $91\times91\times91$. This is significantly larger than the $41\times41$ field utilized in a previous study on denoising 2D ARPES data~\cite{kim2021deep}. We note that beyond the receptive field, the present work is fundamentally distinguished from Ref.~\cite{kim2021deep} by its probabilistic formulation: whereas the previous network provides only a point estimate of the denoised signal, the present one additionally estimates the associated uncertainty, enabling quantitative analysis. The enlarged receptive field allows the network to extract broader contextual information, enabling the successful denoising of extremely noisy data with average voxel values as low as 0.02, as demonstrated on the test data below.

In addition to the denoised signal, the network is trained to generate a corresponding uncertainty map. To achieve this dual output, we optimize the network using a probabilistic loss function, specifically the Laplacian negative log-likelihood (NLL):
\begin{equation*}    
\mathcal{L} = \frac{|y - \hat{y}|}{\sigma/\sqrt{2}} + \log\frac{\sigma}{\sqrt{2}}
\end{equation*}
Here, $y$ is the ground truth, $\hat{y}$ is the predicted denoised intensity, and $\sigma^2$ is the predicted variance, representing the uncertainty. Unlike standard mean squared error (MSE) or mean absolute error (MAE) losses that apply a uniform penalty to all voxels, the NLL dynamically weights the prediction error based on the local uncertainty. In strongly noisy regions, the network can predict a large variance ($\sigma^2$) to reduce the penalty from the first term. To prevent the network from trivially predicting arbitrarily large uncertainty everywhere, the second term acts as a regularizer that penalizes unnecessarily large uncertainties. Mathematically, it is well-established that minimizing this NLL objective function guarantees that the network outputs converge to the approximated mean and variance of the underlying data distribution~\cite{kendall2017uncertainties}.

\begin{figure*}[!t]
	\centering
	\includegraphics[width=1\textwidth]{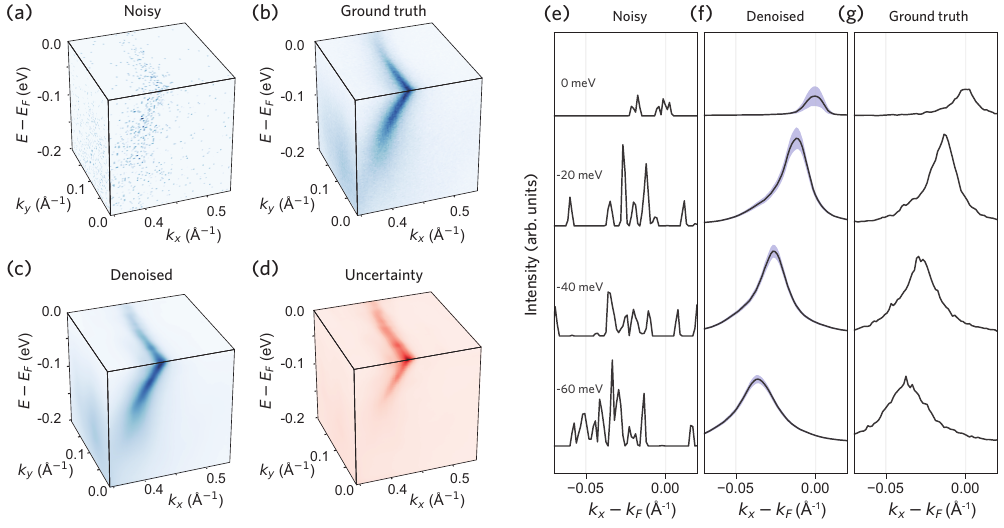}
	\caption{
         {\bf Demonstration of probabilistic denoising on three-dimensional angle-resolved photoemission spectroscopy (ARPES) data.}
        (a) Noisy input data of optimally doped Bi-2212 acquired for 12 seconds.
        (b) Ground truth data acquired over 5 hours. 
        (c) Denoised output.
        (d) Uncertainty map representing the predictive standard deviation.
        (e-g) Momentum distribution curves at $k_y=0$ for the noisy, denoised, and ground truth data, respectively. The blue shaded region in (f) represents the standard deviation ($1\sigma$ uncertainty). The noisy data in (e) are shown after conversion from angle to momentum space; the interpolation involved in the conversion redistributes each detected electron over neighboring grid points, so intensities are in arbitrary units.
	} 

	\label{arpes}
\end{figure*}

\begin{figure*}[!t]
	\centering
	\includegraphics[width=0.95\textwidth]{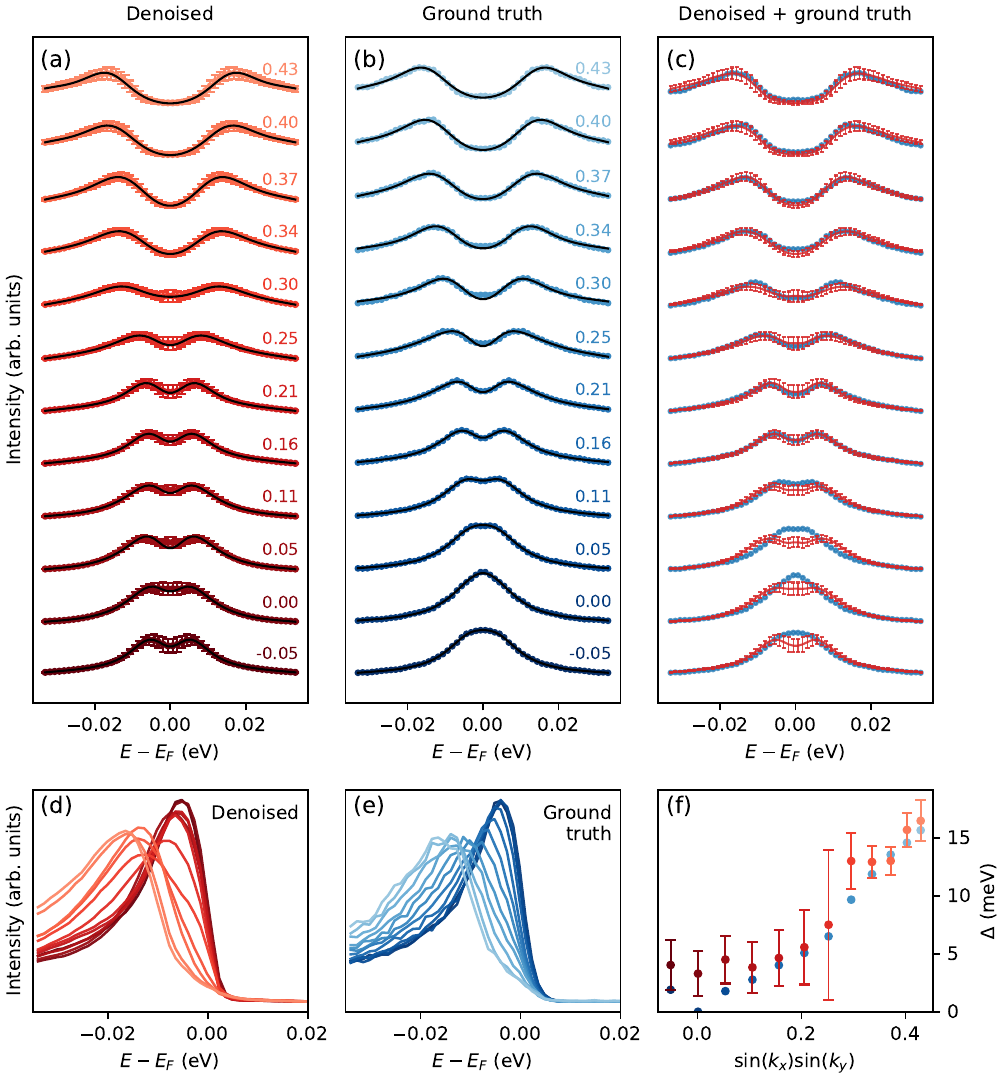}
	\caption{
         {\bf Superconducting gap analysis}
         (a,b) Symmetrized energy distribution curves (colored dots) and the corresponding fits using a phenomenological model (black solid lines) for the denoised and ground-truth data, respectively. The numbers on each energy distribution curve indicate the corresponding $\sin k_x\sin k_y$ values. Error bars represent $1\sigma$ standard deviation of the denoised output.
         (c) Overlay of the plots in (a) and (b).
         (d,e) Energy distribution curves of the denoised and ground truth data, respectively.
         (f) Extracted superconducting gap, where red and blue dots represent the gaps from the denoised and ground-truth data, respectively. Error bars represent $1\sigma$ standard deviation of the fitted parameter. Colors of the dots correspond to the EDCs shown in (a,b).
	} 

	\label{mdc}
\end{figure*}

\begin{figure*}[!t]
	\centering
	\includegraphics[width=1\textwidth]{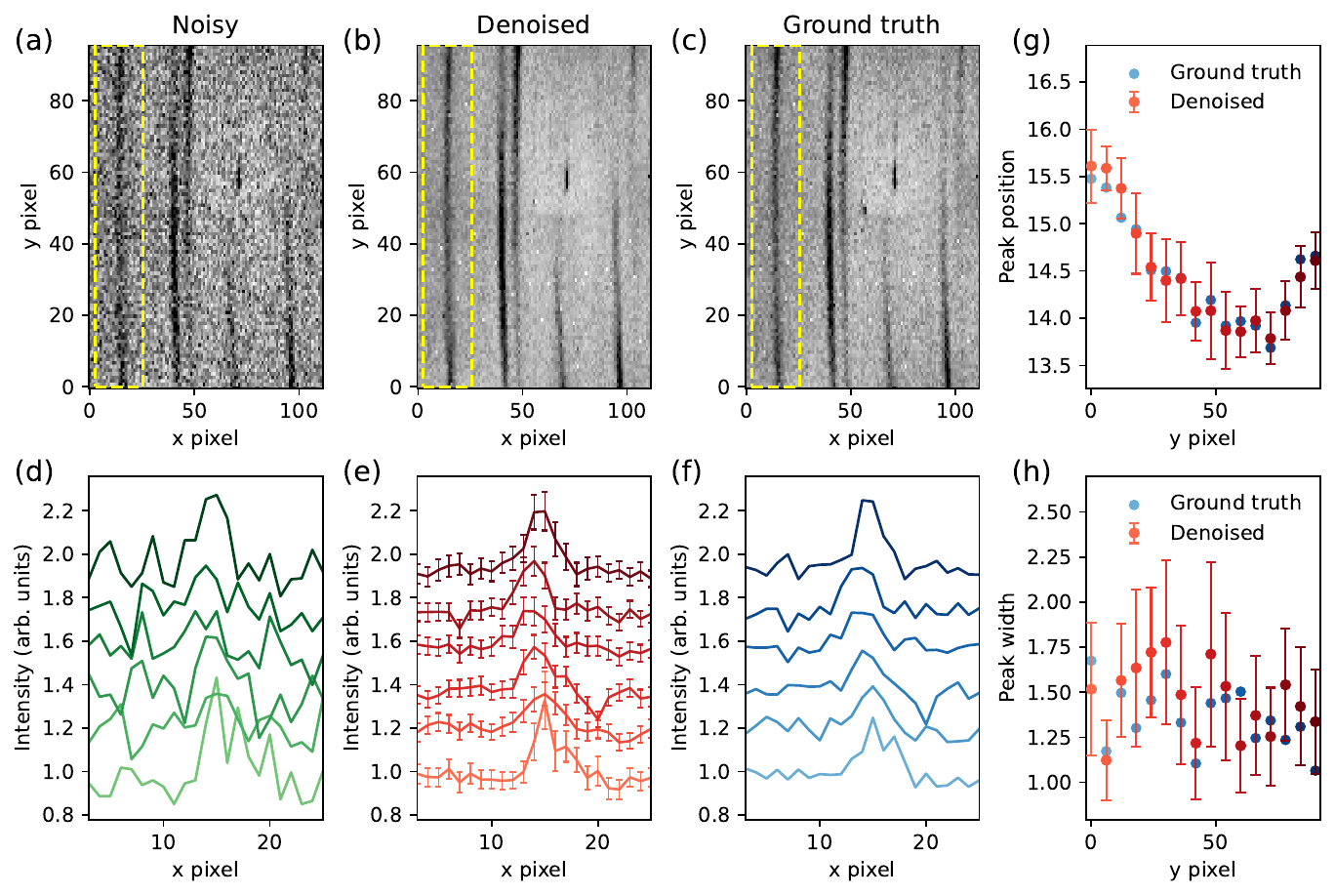}
	\caption{
         {\bf X-ray diffraction (XRD) data analysis}
         (a-c) Noisy, denoised, and ground truth XRD data, respectively. Yellow dashed boxes indicate the regions of interest for peak fitting analyses.
         (d-f) Extracted line spectra along the x-axis from (a-c). Different colors indicate different y pixel value. Error bars in (e) represent $1\sigma$ standard deviation of the denoised output.
         (g,h) Fitted results for the peak position and peak full-width half maximum (FWHM), respectively. Colors of the dots correspond to the line spectra shown in (e) and (f). Error bars in (g,h) represent $1\sigma$ standard deviation of the fitted parameters.
	} 

	\label{XRD}
\end{figure*}
Despite the known effectiveness of this approach, training with the NLL loss requires careful optimization, as it often leads to unstable training and/or underfitting or overfitting~\cite{seitzer2022pitfalls}. This issue becomes critical for 3D data, where the batch size is severely limited by the large data volume. To alleviate this problem, we adopted several stabilization strategies. First, both the predicted output and the ground truth were normalized to unit average, i.e., each 3D volume divided by its mean, before evaluating the loss. This removes the global intensity scale, which carries no physical meaning in spectroscopic measurements, from the optimization and suppresses the associated instability. Second, the total count level was provided as an additional input channel, allowing the network to explicitly optimize its inference based on the count. Finally, relatively strong weight decay of $5 \times 10^{-3}$ was employed to regularize the network parameters, thereby stabilizing the training process and reducing overfitting. We emphasize that the integration of these stabilization techniques was essential in achieving stable network convergence, as training with the NLL loss proved otherwise unstable. Indeed, the training and validation loss curves show stable convergence without signs of overfitting (see Supplemental Material Section~1~\cite{supple}).

To benchmark the performance of the trained network, we used experimental data from Pb-doped Bi$_2$Sr$_2$CaCu$_2$O$_{8+x}$ (Bi-2212). We note that the network was trained on 48 open-source 3D ARPES datasets~\cite{agustsson_2024_12665275} together with 8 datasets measured by us, and the Bi-2212 data serve as an independent test set, excluded from both training and validation (see Methods for details). Figures~\ref{arpes}(a-d) show the noisy input, ground truth, denoised and uncertainty output, respectively. The noisy data was acquired for 12 seconds, with an average count of around 0.02 electrons per voxel over the full 3D volume. Statistically, this represents an extremely sparse condition where approximately 98~\% of the voxels are entirely empty, and only about 2~\% capture a single electron event. At this extremely sparse count level, most voxels contain no detected event, and the network estimate therefore depends more strongly on spectral correlations learned from the training data. In this sense, the output may also be viewed as a reconstruction of the underlying signal. Nevertheless, because the relative Poisson noise increases continuously as the accumulated count decreases, we consider this regime as the extreme low-count limit of the same denoising problem. Under such extreme noise levels, classical noise alleviation methods like Gaussian filtering are completely ineffective~\cite{kim2021deep}. The ground truth data was acquired for 5 hours. Despite the extremely noisy input, the network successfully denoised the data, restoring its main features (see momentum distribution fitting results in Supplemental Material Section~2~\cite{supple}). This aspect is further visualized in the momentum distribution curve (MDC) plots along the nodal line ($k_y=0$) shown in Fig.~\ref{arpes}(e-g). The denoised output recovers a smooth spectrum from data acquired with a three-orders-of-magnitude shorter time than the ground truth, with the accompanying uncertainty map quantifying the reliability of this estimate.

The blue-shaded region in Fig.~\ref{arpes}(f) represents the $1\sigma$ predictive uncertainty. Overall, the predicted uncertainty is proportional to the intensity, which is naturally expected for a Poisson distribution where the standard deviation is $\sqrt{n}$ for an average value of $n$. The statistical calibration of the predicted uncertainties is verified by the normalized residuals, $z=\frac{y-\hat{y}}{\sigma}$. The distribution closely follows the standard normal distribution, and the $1\sigma$ interval of the output covers 64.7~\% of the ground-truth values, in good agreement with the nominal coverage of 68~\% expected for a well-calibrated estimator (see Supplemental Material Section 3 for details~\cite{supple}). This agreement confirms that the network shows no signature of overconfidence, a well-known issue of probabilistic networks trained with NLL objectives~\cite{seitzer2022pitfalls}. Notably, the MDC near the Fermi level exhibits the highest uncertainty despite its relatively weak spectral weight. This strong uncertainty can be attributed to the inherent lack of information required to precisely locate the Fermi edge. Since the Fermi edge is a singular point where the spectral intensity drops abruptly, accurately pinpointing this position with low counts is highly ambiguous. This physical ambiguity thus leads to a high predictive uncertainty in the output of the network.

Beyond merely predicting uncertainty, the key novelty of our approach for scientific research lies in propagating the predicted uncertainty into subsequent numerical analyses, which enables statistical inference, such as hypothesis testing, on the extracted physical parameters. For instance, curve fitting involving the spectral intensity can be performed using the following weighted objective function~\cite{press2007numerical}:
\begin{equation*}
    \chi^2(\vec{\beta_j}) = \sum_{i=1}^{n} w_i [y_i - f(x_i, \vec{\beta_j})]^2 = \sum_{i=1}^{n} \left( \frac{y_i - f(x_i, \vec{\beta_j})}{\sigma_i} \right)^2
\end{equation*}
where $y_i$ is the denoised intensity, $f(x_i, \vec{\beta_j})$ is the theoretical model with a set of fitting parameters $\vec{\beta_j}$, and $\sigma_i$ is the voxel-wise uncertainty generated by our model. By incorporating $\sigma_i$ as the statistical weight, the uncertainty of the neural network's output is transferred to the physical parameters extracted from the spectra. The covariance matrix $C$ of the optimized parameters can then be derived from the  Hessian of $\chi^2$:
\begin{equation*}    
C_{ij} = 2 \left( \frac{\partial^2 \chi^2}{\partial \beta_i \partial \beta_j} \right)^{-1}
\end{equation*}
The $1\sigma$ standard uncertainty for the $j$-th fitted parameter, $\Delta \beta_j$, is given by the square root of the corresponding diagonal element:$$\Delta \beta_j = \sqrt{C_{jj}}$$

However, the uncertainty estimation assumes that the voxel-wise uncertainties are statistically independent. In practice, however, the denoised spectra exhibit a strong correlation between adjacent data points, as the smooth denoised output cannot vary independently from voxel to voxel, which leads to an underestimation of the uncertainty of the fitted parameters. To compensate for this, the standard uncertainty of the fitted parameters must be adjusted using the effective degrees of freedom ($N_{eff}$). The effective degrees of freedom can be estimated by determining the correlation length through the autocorrelation of the spectrum. By computing the integrated autocorrelation length $\lambda$, the effective degrees of freedom are reduced to $N_{eff} = N / \lambda$, where $N$ is the original number of voxels in a spectrum. We note that $\lambda=1$ for uncorrelated data, and $\lambda>1$ for correlated data. Taking into account this factor, the corrected covariance matrix and $1\sigma$ standard uncertainty are given by:

\begin{equation*}
C_{corrected} = C \left( \frac{N}{N_{eff}} \right) = \lambda C
\end{equation*}
\begin{equation*}
\Delta \beta_{j}^{corrected} = \sqrt{(C_{corrected})_{jj}} = \Delta \beta_j \sqrt{\lambda}
\end{equation*}

Having established the framework for analyzing the probabilistic output, we now demonstrate the superconducting gap analysis results obtained from the energy distribution curve (EDC) fittings. Figure 3 shows the superconducting gap fitting results for the Bi-2212 data. Figures 3(a) and 3(b) display the symmetrized EDCs of the denoised and ground truth data, respectively, at the Fermi momenta across different momentum points. We note that the denoised data is generated from the same experimental data shown in Fig.~2, which is acquired for 12 seconds. The symmetrized EDCs were fitted with the phenomenological model~\cite{norman1998phenomenology}, based on the self-energy $\Sigma (\omega)=-i\Gamma_1+\Delta^2/\omega$, where $\Delta$ is the superconducting gap and $\Gamma_1$ is the single-particle scattering rate. We note that this model is widely used for extracting the superconducting gap from symmetrized EDCs~\cite{wu2024nodal}. In the ground truth data, the symmetrized EDCs show the absence of the gap near the nodal point and a gap opening at off-nodal points. While the EDCs of the denoised data agree well at off-nodal points, the spectral shape deviates slightly near the nodal point. 

The gap sizes extracted from the fitting are summarized in Fig. 3(f). Since the sample is aligned along the nodal direction, we use $\sin k_x\sin k_y$ as the $x$-axis, against which the $d$-wave superconducting gap is expected to exhibit linear dependence. As expected, the ground truth data exhibit this linear behavior. Surprisingly, the denoised data acquired only for 12 seconds roughly follow the ground truth, except near the nodal region, where the model tends to overestimate the gap size.

The propagated uncertainty allows this discrepancy to be assessed quantitatively. The fitted nodal gap is $\Delta = 3.3 \pm 1.9 ~\text{meV}$. While its $1\sigma$ interval does not include zero, the 95~\% confidence interval (approximately $2\sigma$) does, so the null hypothesis ($\Delta=0$) cannot be rejected at the standard 95~\% confidence level. The apparent nodal gap is therefore not statistically significant. This illustrates the practical value of uncertainty-aware denoising, since a naive point estimate would report a false gap indistinguishable from a genuine one, whereas the propagated uncertainty correctly identifies the nodal value as statistically consistent with zero.

This discrepancy near the nodal point indicates that the impact of limited information is most pronounced in the nodal region, despite its high spectral weight. While the nodal point possesses a sharp and intense Fermi edge, this feature makes the gap extraction highly susceptible to the precision of the predicted Fermi level. In the absence of sufficient information from short-acquisition time, even a minor uncertainty in the Fermi level significantly distorts the sharp spectral profile during symmetrization, manifested as a false gap opening. The nodal region, in turn, becomes highly ambiguous for the model under noisy conditions. This interpretation is further supported by the high uncertainty predicted near the nodal points. To verify the generality of the trained network, we applied it to additional ARPES data from a Bi$_2$Te$_3$ thin film, where it achieves comparable denoising performance (see Supplemental Material Section 4~\cite{supple}).

To demonstrate the broad applicability of our approach beyond spectroscopy, we applied the method to 2D XRD data, as XRD is a widely used technique across a broad range of disciplines in science~\cite{cullity1957elements}. For this demonstration, we adopted data and source code from an open-source repository~\cite{oppliger2024weak,oppliger2022x,oppliger_zenodo_10245374}. Using these, the network is trained separately from that for ARPES data. Based on the original code, we slightly modified the network architecture to generate a dual output by changing the output channels to two and adopting the Laplacian NLL as the loss function. The XRD images shown in Fig.~\ref{XRD}(a-c) demonstrate that the modified network effectively removes noise from the raw data.

Following the qualitative confirmation of noise reduction, we extracted line spectra (Fig.~\ref{XRD}(d-f)) for a quantitative evaluation. Similar to our previous analysis, we performed single Gaussian peak fittings on these extracted profiles to determine the peak positions (Fig.~\ref{XRD}(g)) and peak widths (Fig.~\ref{XRD}(h)) with the aforementioned uncertainty propagation method. The peak positions and widths extracted from the denoised data agree with the ground truth, and the error bars are successfully propagated. This successful demonstration confirms that our approach is highly versatile, proving its applicability across diverse scientific measurements regardless of the experimental techniques or dimensionality of the data.

\begin{figure}[!t]
	\centering
	\includegraphics[width=0.45\textwidth]{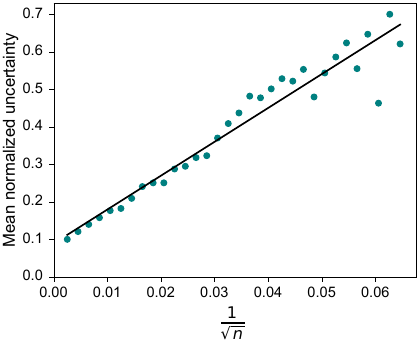}
	\caption{
         {\bf Scaling of uncertainty as a function of total count $n$.}
         Each data point is randomly sampled from the ground truth data to achieve the target total counts.
	} 

	\label{Fig1}
\end{figure}

\section*{Discussion}

From the perspective of information theory, the total Fisher information contained in a spectrum with a total count $n$ is proportional to $n$, assuming each count represents an independent detection event~\cite{kay1993statistical}. We note that this independence is a property of the counting event itself and is distinct from the correlation of the denoised output discussed above, which reflects the smoothness of the underlying spectrum. According to the Cramér-Rao bound, the variance of any unbiased estimator cannot be lower than the inverse of the total Fisher information:
\begin{equation*}
Var(\hat{\theta}) \geq \frac{1}{I_n(\theta)} = \frac{1}{n \cdot I_1(\theta)}
\end{equation*}
where $I_n(\theta)$ is the total Fisher information for a parameter $\theta$ and $I_1(\theta)$ is Fisher information of single electron count~\cite{cover1999elements}. Accordingly, the fundamental lower bound for the uncertainty (standard deviation) scales as $1/\sqrt{n}$. Figure 5 shows the mean normalized uncertainty as a function of $1/\sqrt{n}$, respectively. It shows a clear linear dependence on $1/\sqrt{n}$, indicating that the trained network extracts information with near-optimal efficiency. In other words, the probabilistic denoiser operates close to the fundamental information limit imposed by counting statistics.

This scaling highlights an important experimental implication. Since the uncertainty scales as $1/\sqrt{n}$, the marginal gain in precision decreases rapidly with increasing acquisition time. That is, a large fraction of statistically useful information is obtained during the early stages of measurement. In practical spectroscopic experiments, this means that extending the measurement time far beyond the initial acquisition therefore yields only marginal gains in statistical precision and is often performed primarily to obtain visually smoother spectra. 

Instead, a more efficient strategy is to combine short acquisition measurements with denoising, allowing reliable extraction of the underlying spectral features while simultaneously quantifying the uncertainty. In this sense, probabilistic denoising enables deep learning to act not merely as a visualization tool, but as a quantitatively reliable method for extracting physical information from noisy experimental data. By doing so, the time required for data acquisition can be drastically reduced, which has been a fundamental bottleneck in spectroscopic experiments. 

Furthermore, our approach is not limited to the demonstration shown here. While we have focused on ARPES and XRD, this probabilistic denoising strategy is broadly applicable to any multidimensional measurement where acquisition time is a critical bottleneck. We note that the approach can be extended to even higher-dimensional data, where more contextual information is expected to yield better denoising performance once properly trained. Building upon recent efforts to integrate machine learning with scientific data analysis~\cite{peng2020super,kim2021deep,oppliger2024weak,meyer2025line,agustsson2025autoencoder,majchrzak2025machine,chen2025detecting,yoon2023deep,joucken2022denoising,xian2023machine,liu2023removing,ekahana2023transfer,na2025simulation,ziatdinov2022atomai}, our framework provides a crucial advancement: it establishes a rigorous pathway for deploying deep learning not just as a visualization aid, but as a quantitative tool for physical analysis. The code and the trained ARPES network are publicly available, both to ensure reproducibility and to allow direct application to 3D ARPES data.

\section*{Methods}

\textit{Network architecture and training}---The training data set consists of 56 high-count 3D ARPES data, randomly split into 51 for training and 5 for validation. Part of the data set was sourced from an open-source data repository~\cite{agustsson2025autoencoder,agustsson_2024_12665275}. The data were resized into a $128\times128\times128$ grid with a sum-conserving resizing method and augmented eightfold by flips and rotations along the $k_x$ and $k_y$ axes. During training, noisy inputs were generated on the fly from high-count data in real time. The average count per voxel ranges from 0.005 to 0.1 with a logarithmic distribution. We note that, although the number of datasets is modest, each is a full 3D volume: the training targets amount to be $51 \times 128^3 \approx 1.1 \times 10^8$ voxels, far exceeding the total number of trainable parameters in the network ($5.6 \times 10^6$). Together with the limited feature diversity of ARPES spectra and the augmentation strategies described above, this dataset is proven to be sufficient for training, as confirmed by the validation loss closely following the training loss without any overfitting signature (see Supplemental Material Section 1~\cite{supple}). After each convolutional layer, the data were fed into a group normalization (GN) layer and a sigmoid linear unit (SiLU) activation function. We optimized the network using AdamW~\cite{loshchilov2017decoupled} with a relatively high weight decay of $5\times10^{-3}$ to prevent overfitting. The learning rate followed a cosine decay schedule with a peak value of $5 \times 10^{-4}$, preceded by a warm-up phase in which the rate was increased linearly from zero to the peak over the first 5 epochs. Due to GPU memory (VRAM) constraints, the batch size was limited to 1. The network was trained using a NVIDIA RTX 2060 with the PyTorch framework.

\textit{ARPES experiments}—For the demonstrative data shown in Figs. 2, 3, laser-based ARPES measurements were performed using a lab-based system at Seoul National University. A home-built Yb-doped fiber laser with a center wavelength of 1064 nm, a pulse width of 50 ps and a repetition rate of 1.5 MHz was used to generate 7 eV (177 nm) light. Spectra were acquired using a Scienta ARTOF 10k electron analyzer with p-polarized light. The energy resolution was set to better than 2 meV.

\section*{Data availability}
The experimental data that support the findings of this article are publicly available~\cite{kim_data2026}.

\section*{Code availability}
The code used for training the neural networks, together with the trained network, is publicly available from Ref.~\cite{kim_code2026}.

%

\section*{Acknowledgments}
We thank Dongjoon Song for providing high-quality Bi-2212 samples; Sangjae Lee, Jae Hyuck Lee, Suyoung Lee, Saegyeol Jung and Jongkeun Jung for providing the 3D ARPES data used in the training dataset; and Youngdo Kim for helpful comments on the superconducting gap analysis. This research was also supported by the National Research Foundation of Korea (NRF) grant funded by the Korean government (MSIT) (No. 2022R1A3B1077234). It was also supported by GRDC(Global Research Development Center) Cooperative Hub Program through the National Research Foundation of Korea(NRF) funded by the Ministry of Science and ICT(MSIT) (RS-2023-00258359).

\section*{Author contributions}
Y.K. and C.K. conceived and designed the project. Y.K. developed the deep learning algorithm, wrote the software, and performed the ARPES measurements. Y.K. and C.K. wrote the manuscript. Both authors discussed the results and reviewed the manuscript.

\end{document}


\title{Supplemental Material for \\ Probabilistic denoising for reliable signal extraction in spectroscopy}

\newcommand{\red}[1]{{\textcolor{red}{#1}}}

\date{\today}

\maketitle
\tableofcontents
\newpage
\section{Network architecture and training details}
\begin{figure*}[h]
	\centering
	\includegraphics[width=0.43\textwidth]{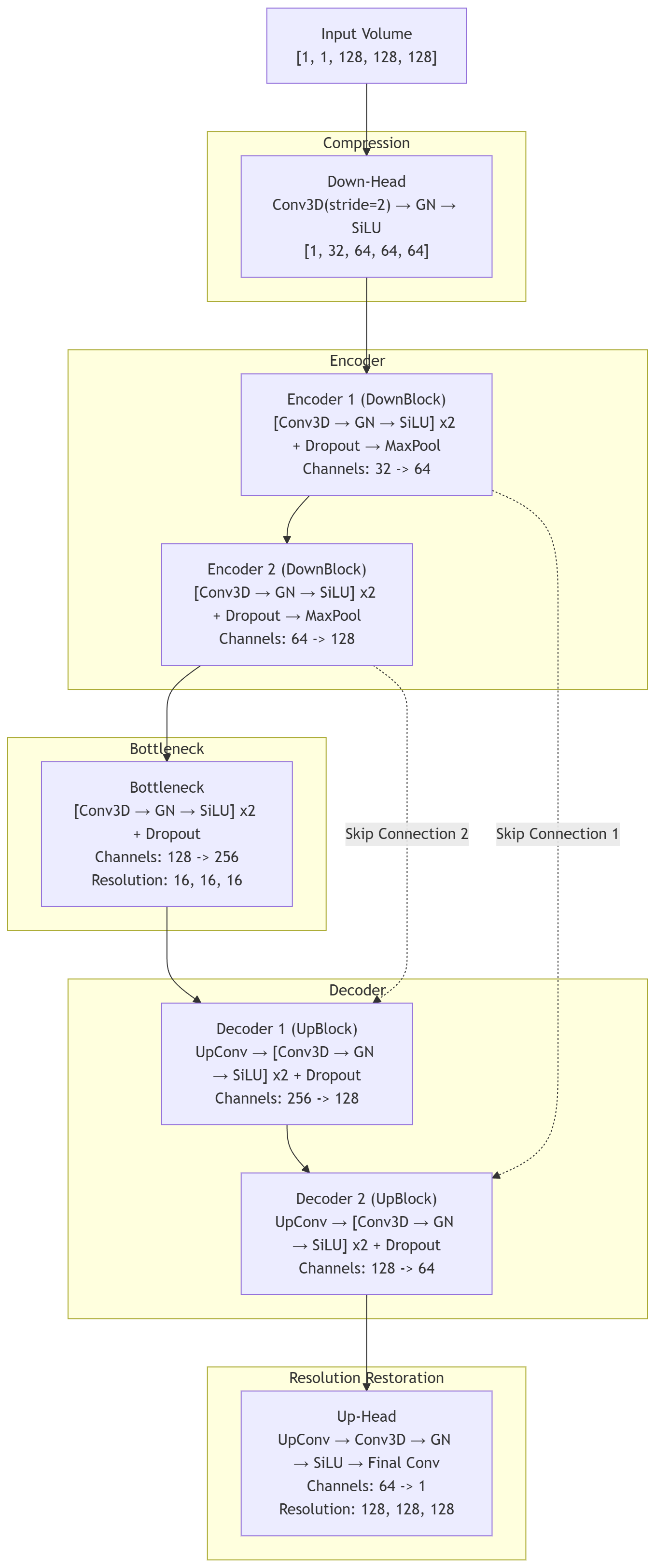}
	\caption{
         {\bf Detailed network structure.}
         }
	\label{fig_network}
\end{figure*}

As specified in the main manuscript, the network utilizes an autoencoder structure (Fig.~\ref{fig_network}). The architecture follows a U-Net design, but with the skip connections turned off, which reduces it to an autoencoder structure. We compared the two configurations and found that the network without skip connections yields better denoising performance. We attribute this to the nature of the task: the input is dominated by high-frequency Poisson noise while the target output is a smooth spectrum, so the skip connections mainly pass noise components of the input directly to the output, bypassing the information bottleneck of the autoencoder. Removing them forces all information to pass through the bottleneck, which suppresses the noise transfer.

\begin{figure*}[h]
	\centering
	\includegraphics[width=0.6\textwidth]{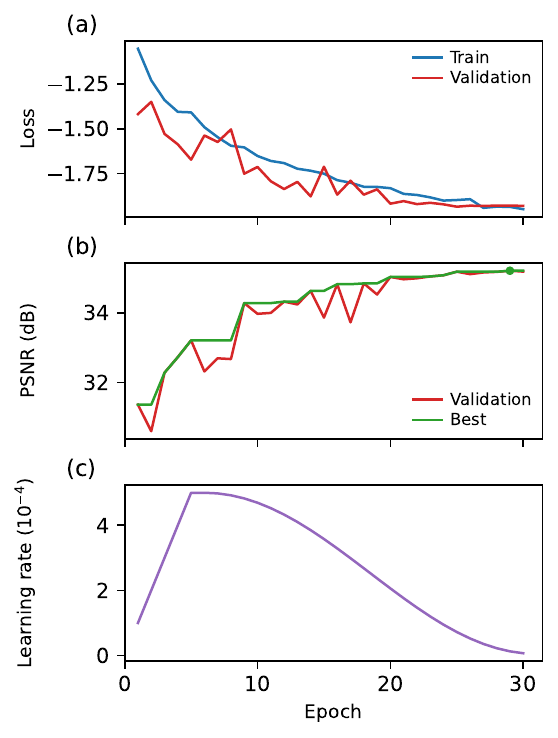}
	\caption{
         {\bf Training curves.}
          (a) Training and validation loss. 
          (b) Validation peak signal-to-noise ratio (PSNR).
          (c) Learning rate as a function of epoch.
         }
	\label{fig_training}
\end{figure*}

Training curves are shown in Fig.~\ref{fig_training}. The training and validation losses remain comparable, and the validation loss overall decreases monotonically over the epochs (Fig.~\ref{fig_training}(a)), indicating that the network is not overfitted. For stable convergence, the learning rate follows a cosine annealing schedule preceded by a linear warm-up phase, in which the rate is increased from zero to its peak value of $5 \times 10^{-4}$ over the first 5 epochs (Fig.~\ref{fig_training}(c)).

\newpage
\section{Momentum distribution curve analysis}

\begin{figure*}[h]
	\centering
	\includegraphics[width=0.7\textwidth]{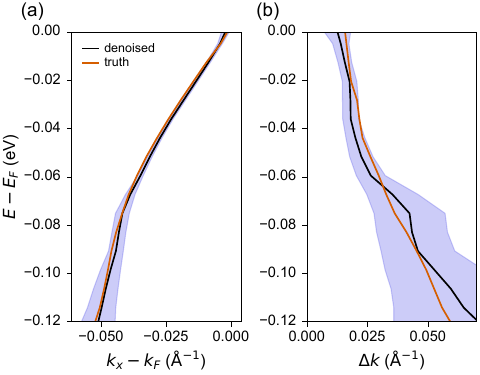}
	\caption{
         {\bf Momentum distribution curve fitting analysis.}
         (a) Fitted peak positions from the denoised data (black solid line, with the blue shaded area representing the $1\sigma$ uncertainty) and the ground truth data (brown solid line). (b) Fitted peak widths, shown in the same manner as in (a).
         }
	\label{fig_mdc}
\end{figure*}

To quantitatively evaluate the denoised output of the Bi-2212 data shown in Fig.~2 of the main text, we performed momentum distribution curve (MDC) fitting along the nodal direction. Each MDC was fitted with a Lorentzian function, with the voxel-wise predictive uncertainties used as statistical weights following the propagation formalism described in the main text. Figure~\ref{fig_mdc} shows the fitted peak positions and widths as a function of binding energy, with the ground-truth results overlaid for direct comparison. The peak positions from the denoised data agree with the ground truth within the $1\sigma$ uncertainty over the measured energy range, and the peak widths show comparable agreement. This confirms that the denoised output reproduces not only the band dispersion but also the linewidth, with uncertainties that correctly account for the deviation from the ground truth.

\newpage
\section{Statistical calibration of the predicted uncertainties}
To verify that the predicted uncertainties represent the actual deviation from the ground truth, we investigate the distribution of the normalized residuals of the Bi-2212 data shown in Figs.~2 and 3,
\begin{equation*}
z_i = \frac{y_i- \hat{y}_i}{\sigma_i},
\end{equation*}
where $y_i$ and $\hat{y}_i$ are the voxel-wise values of the ground truth and the denoised output, respectively. Since the ground
truth itself is a measurement with finite counts, its statistical fluctuation also contributes to the residuals. The denominator therefore combines the two contributions,
\begin{equation*}
\sigma_i = \sqrt{\sigma_{\mathrm{model},i}^{2} +
\sigma_{\mathrm{truth},i}^{2}},
\end{equation*}
where $\sigma_{\mathrm{model}}$ is the predictive uncertainty of the network and $\sigma_{\mathrm{truth}}$ is the Poisson noise of the
ground-truth data. To restrict the evaluation to regions containing actual spectral weight, we include only voxels with nonzero ground-truth counts ($y_i > 0$).

\begin{figure*}[h]
	\centering
	\includegraphics[width=0.55\textwidth]{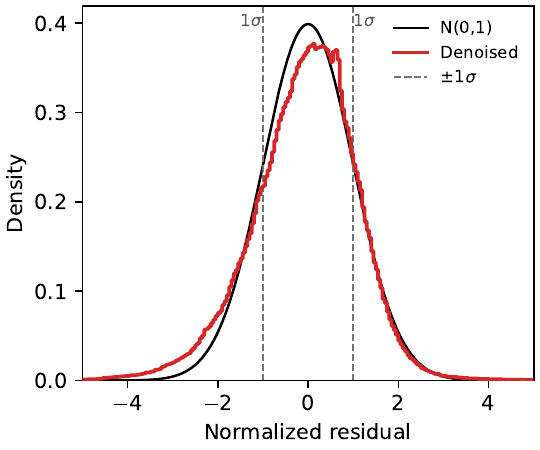}
	\caption{
         {\bf Histogram of the normalized residuals of the Bi-2212 data.}
         The residual is obtained from $z_i = \frac{y_i- \hat{y}_i}{\sigma_i}$, where $y_i$ and $\hat{y}_i$ are the voxel-wise values of the ground truth and the denoised output, respectively. $\sigma_i$ is the calibrated standard deviation.
         }
	\label{fig_residual}
\end{figure*}

Figure~\ref{fig_residual} shows the histogram of $z$ over the valid 3D volume, which closely follows the standard normal distribution $N(0,1)$. Quantitatively, 64.7~\% of the ground-truth values fall within the predicted $1\sigma$ interval, in good agreement with the nominal coverage of 68~\%. This confirms that the network is statistically well
calibrated and shows no signature of overconfidence, a well-known failure mode of probabilistic networks trained with an NLL loss function.

\newpage
\section{Denoising of Bi$_2$Te$_3$ thin film data}

\begin{figure*}[h]
	\centering
	\includegraphics[width=1\textwidth]{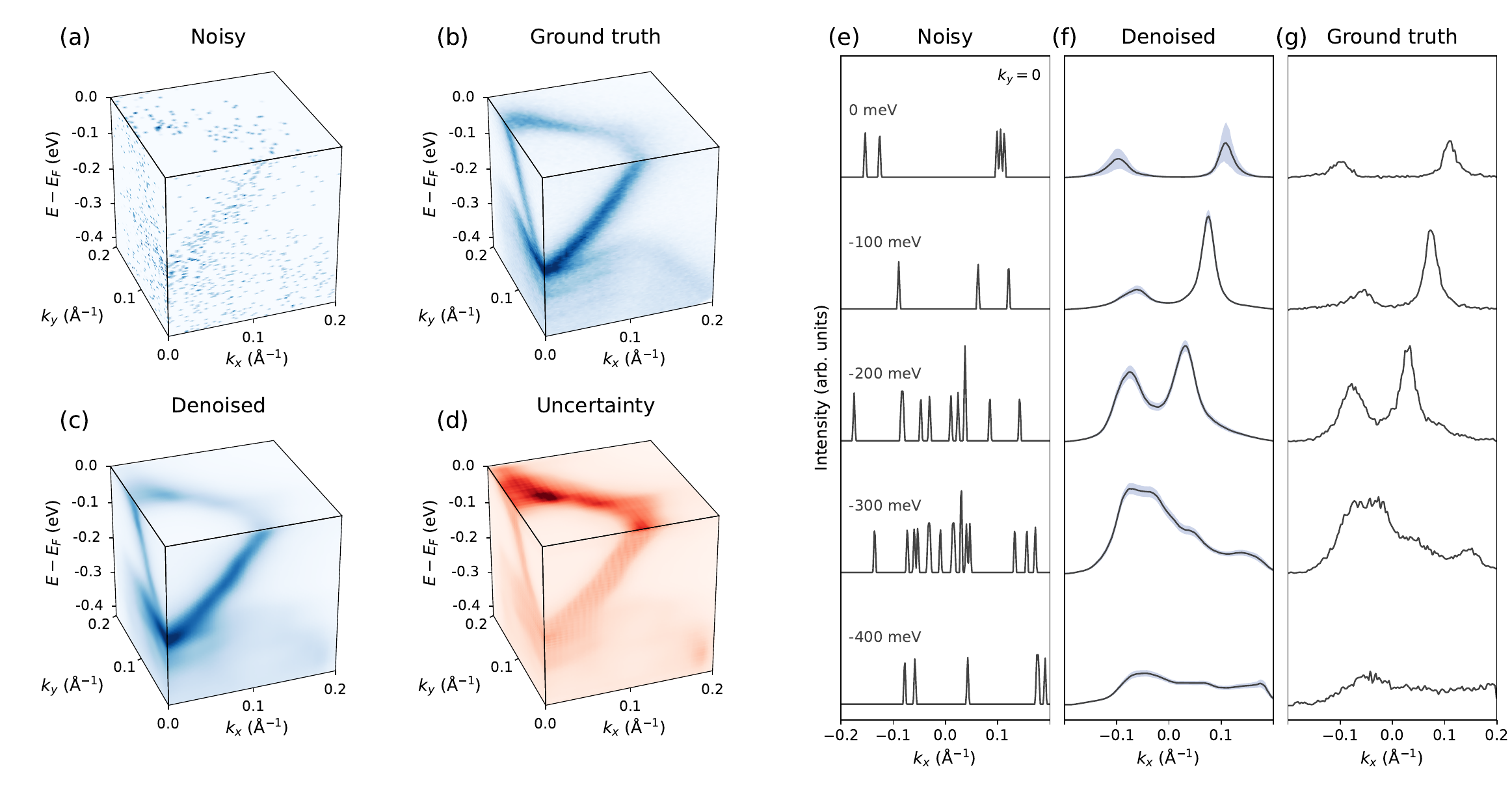}
	\caption{
        \textbf{Denoising results on Bi$_2$Te$_3$ thin film.} (a) Noisy input data with an average voxel count of 0.02. (b) Ground truth data. (c) Denoised output. (d) Uncertainty map representing the predictive standard deviation. (e-g) Momentum distribution curves at ky = 0 for the noisy, denoised, and ground truth data, respectively. The blue shaded region in (f) represents the standard deviation ($1\sigma$ uncertainty).
         }
	\label{fig_bi2te3}
\end{figure*}

As an additional test of the trained network, we applied it to ARPES data from a Bi$_2$Te$_3$ thin film, which was not included in the training or validation set. The noisy input is a direct experimental measurement with an average count of 0.02 electrons per voxel. Figure~\ref{fig_bi2te3} shows the noisy input, the ground truth, the denoised output, and the
predictive uncertainty map, together with the momentum distribution curves at $k_y = 0$. Despite the distinct spectral features of the sample, the network successfully denoises the data, restoring the main spectral features, and the predicted uncertainty map exhibits the same qualitative behavior as in the Bi-2212 case. The uncertainty is overall proportional to the intensity, while the largest uncertainty appears near the Fermi level, reflecting the same difficulty of pinpointing the sharp Fermi edge from sparse counts. This demonstrates that the trained network is not specific to the Bi-2212 data but applies to ARPES spectra of different samples.

